\documentclass[aps,prb,twocolumn,showpacs,superscriptaddress]{revtex4}
\usepackage{graphicx}
\usepackage{amssymb,amsmath}
\usepackage{color}

\begin{document}

\title{Magnetotransport through graphene nanoribbons at high magnetic fields}

\author{S.~Minke}
\thanks{n\'{e}e S.~Schmidmeier.}
\affiliation{Institute of Experimental and Applied Physics, University of Regensburg, 93040 Regensburg, Germany}

\author{S.~H.~Jhang}
\affiliation{Institute of Experimental and Applied Physics, University of Regensburg, 93040 Regensburg, Germany}

\author{J.~Wurm}
\affiliation{Institute of Theoretical Physics, University of Regensburg, 93040 Regensburg, Germany}

\author{Y.~Skourski}
\affiliation{Dresden High Magnetic Field Laboratory,
Helmholtz-Zentrum Dresden-Rossendorf, 01314 Dresden, Germany}

\author{J.~Wosnitza}
\affiliation{Dresden High Magnetic Field Laboratory,
Helmholtz-Zentrum Dresden-Rossendorf, 01314 Dresden, Germany}

\author{C.~Strunk}
\affiliation{Institute of Experimental and Applied Physics, University of Regensburg, 93040 Regensburg, Germany}

\author{D.~Weiss}
\affiliation{Institute of Experimental and Applied Physics, University of Regensburg, 93040 Regensburg, Germany}

\author{K.~Richter}
\affiliation{Institute of Theoretical Physics, University of Regensburg, 93040 Regensburg, Germany}

\author{J.~Eroms}
\email{jonathan.eroms@physik.uni-regensburg.de}
\affiliation{Institute of Experimental and Applied Physics, University of Regensburg, 93040 Regensburg, Germany}

\begin{abstract}
We have investigated the magnetoresistance of lithographically prepared single-layer graphene nanoribbons in pulsed, perpendicular magnetic fields up to 60~T and performed corresponding transport simulations using a tight-binding model and several types of disorder. In experiment, at high carrier densities we observe Shubnikov-de Haas oscillations and the quantum Hall effect, while at low densities the oscillations disappear and an initially negative magnetoresistance becomes strongly positive at high magnetic fields. The strong resistance increase at very high fields and low carrier densities is tentatively ascribed to a field-induced insulating state in the bulk graphene leads. Comparing numerical results and experiment, we demonstrate that at least edge disorder and bulk short-range impurities are important in our samples.
\end{abstract}

\pacs{72.80.Vp, 73.43.Qt, 73.22.Pr}

\maketitle

\section{Introduction}
For the application of graphene in nanoelectronics one has to understand the behavior of graphene nanostructures, in particular graphene nanoribbons (GNRs). They were theoretically predicted to show either metallic or insulating behavior around the charge neutrality point, depending on their crystallographic orientation. In experiment, however, GNRs always exhibit an insulating state close to the charge neutrality point (CNP)~\cite{Han2007}, which is dominated by disorder rather than a confinement-induced gap in the spectrum~\cite{Stampfer2009,Gallagher2010}. A clear proof of conductance quantization only appeared very recently in ultra-clean suspended nanoribbons \cite{Tombros2011}. Furthermore, in clean zigzag edges, a magnetic state has been predicted \cite{Fujita1996,Louie2006}, but so far it has remained elusive in transport experiments. At present, therefore, the behavior of GNRs is mainly governed by extrinsic defects rather than their intrinsic properties, and information on the nature of those defects is highly desired.

In previous experiments, large disorder was attributed to cause strong localization effects which influence the magnetoconductance \cite{Oostinga2010}. Poumirol \textit{et al.} report a large positive magnetoconductance and explain this by simulations which take into account different types of disorder. They affirm the qualitative behaviour, but the computed conductance remains larger than the experimental ones. Also, an unambiguous separation of bulk and edge disorder was not possible~\cite{Poumirol2010}.
Here, we present magnetotransport measurements on GNRs in magnetic fields of up to 60~T and corresponding tight-binding simulations with several types of realistic bulk and edge disorder. By considering the magnetoconductance close to the Dirac point and at high densities, we observe characteristic signatures of bulk and edge disorder and can disentangle their contributions to transport in GNRs.

\section{Experimental Details}

\begin{figure}
\includegraphics[width=8.4cm]{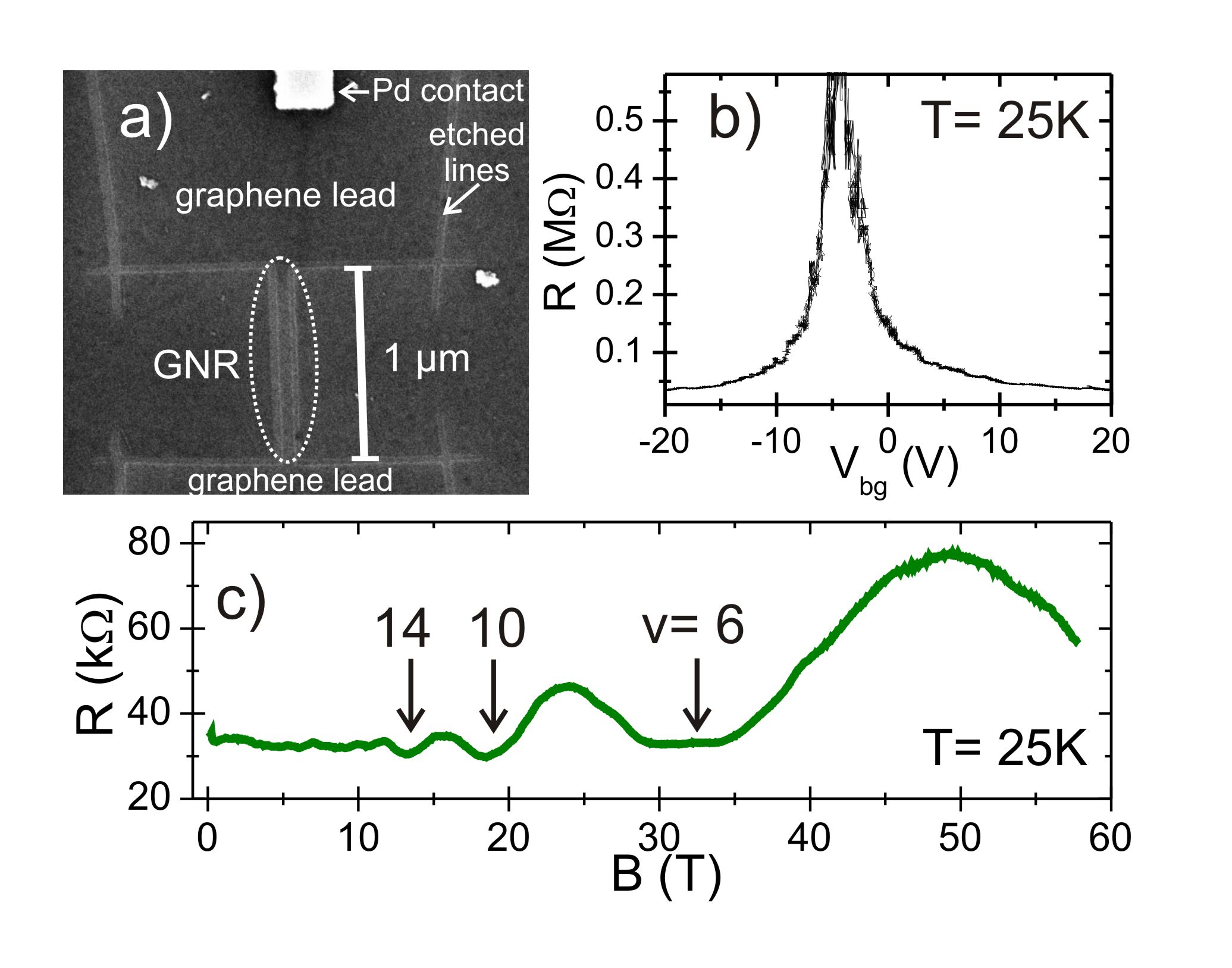}
\caption[GNR]{(Color online) (a) Scanning electron microscope image of a typical sample. The length of the GNRs is 1~$\mu$m, the width 70~nm. In the upper part of the image a palladium contact is visible. (b) Two-terminal resistance as a function of $V_{bg}$ at $T$=~25~K and zero magnetic field. (c) Magnetoresistance trace at $V_{bg}=-20$~V, showing quantum Hall features at $\nu=6,10$ and 14.}
\label{GNR}
\end{figure}

Single-layer graphene is deposited on a highly doped silicon wafer with a 300~nm thick SiO$_{2}$ layer by conventional exfoliation. The graphene nanoribbons were defined by electron-beam lithography and oxygen plasma reactive ion etching. For the transport measurements, palladium contacts were attached to the GNRs. A scanning electron micrograph of the sample discussed here is shown in Fig.~\ref{GNR}(a). The DC magnetotransport measurements with 10~mV DC bias were done in pulsed perpendicular magnetic fields at temperatures between 1.8 and 125~K. Typical pulse durations were ranging from 100 to 300 ms. During the pulse the current through the GNR was converted to a voltage signal by a current-to-voltage amplifier and recorded by a high-speed oscilloscope and data recorder. In total two single-layer nanoribbons have been measured which show similar behavior. Here, we focus on data from one device. Figure \ref{GNR}(b) shows the resistance $R$ of the nanoribbon as a function of back-gate voltage $V_{bg}$ at $T$= 25~K and zero magnetic field. The sharp peak at $V_{bg}=V_{CNP}=-4.4$~V indicates the charge neutrality point. After patterning, the hole mobility $\mu$ of the ribbons is about 590~cm$^{2}$/Vs at $V_{bg}$=~-15~V~\cite{FN3}.
Figure \ref{GNR}(c) shows a magnetoresistance curve taken at high carrier density~\cite{FN1}. A quantum Hall plateau at $\nu=6$~\cite{FN2} and Shubnikov-de Haas oscillations for $\nu=$~10 and 14 are observed. Signatures of Hall states were already found in previous experiment~\cite{Ribeiro2011}. From the zero-field mobility and the condition $\mu B \gg 1$ we would not expect to observe quantum Hall features at $\nu=14$, at 13~T. This is already an indication that the high field changes the impact of disorder on transport in our sample.

\section{Density and Temperature Dependence}

\begin{figure}
\includegraphics[width=8.4cm]{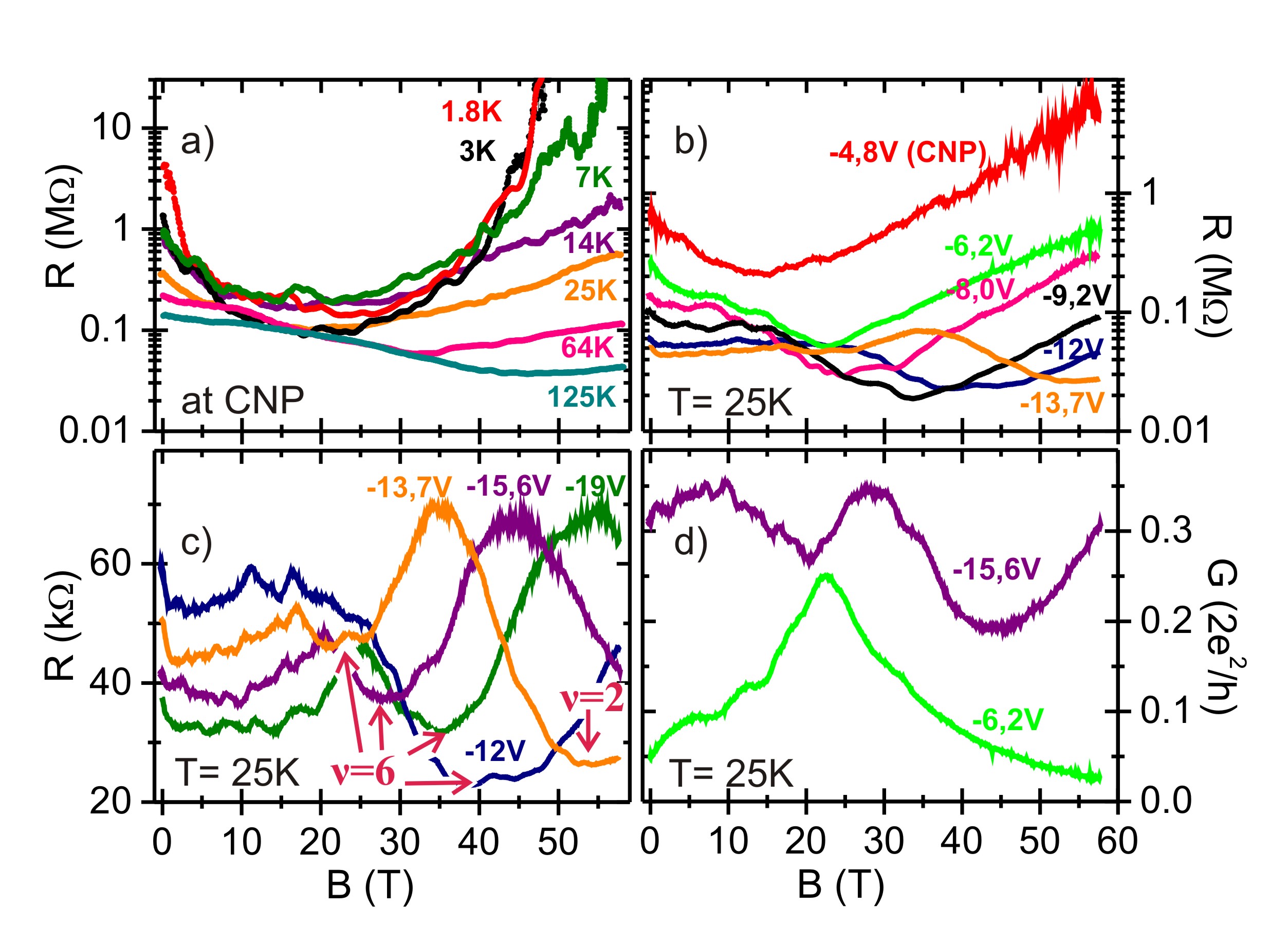}
\caption[alle2]{(Color) (a) Magnetoresistance of the GNR for various temperatures at the charge neutrality point. (b) Magnetoresistance for different gate voltages close to the CNP and (c) further away from the CNP at $T$=~25~K. The arrows and the numbers indicate the corresponding filling factors $\nu$ of the quantum Hall state, $\nu$=~2 and 6. (d) Conductance as a function of magnetic field for $V_{bg}$=~-15.6 and -6.2~V.}
\label{alle2}
\end{figure}

Let us now consider the density and temperature dependence of the magnetoresistance  in more detail. First, we will focus on the transport properties at gate voltages close to the CNP. For all temperatures we tuned the backgate voltage such that the samples remained as close as possible to the CNP. In Fig.~\ref{alle2}(a), the magnetoresistance is plotted for various temperatures ranging from 1.8 to 125~K. For all temperatures a resistance decrease is observed for fields up to about 20~T, so that the ribbon crosses over from a highly resistive state to a metallic regime. Subsequently, it is followed by a prominent resistance increase. The divergent form of the latter increase suggests that the nanoribbon approaches a field-induced insulating state.

In order to better comprehend the observed behavior, we studied the magnetoresistance for different gate voltages ranging from -4.8 to -13.7~V at $T$=~25~K. As one can see in Fig.~\ref{alle2}(b), the observed divergence of the resistance at very high fields only appears for gate-voltages close to the CNP ($|V_{bg}-V_{CNP}|<$~9~V).
At higher densities [see Fig.~\ref{alle2}(c)], we observe weak localization at fields up to 1~T, a fairly constant resistance up to about 20~T, and then pronounced resistance oscillations.
These oscillations can be identified as Shubnikov-de Haas (SdH) oscillations, which can be assigned to Hall-plateau values of single-layer graphene ($\nu=2$ and 6). The capacitive coupling $C_{g}$ of the nanoribbon to the back-gate, which strongly depends on the ribbon dimensions, was calculated using a finite-element model, yielding $C_{g}=576$~aF/$\mu$m$^{2}$ for a 70~nm wide GNR. Plotting the fan diagram of the minima of the SdH oscillations gives a coupling $C_{g}$ of 560~aF/$\mu$m$^{2}$, which matches the calculated value well. Therefore, the carrier density is estimated as $n \approx 3.5\times 10^{15}~\mathrm{m}^{-2}~\times (V_{bg}-V_{CNP})$ and the Fermi-energy scales as $E_{F}\approx 69\ \mathrm{meV} \times \sqrt{|V_{bg}-V_{CNP}|}$, where $V_{bg}$ and $V_{CNP}$ are given in Volts.

For easier comparison to the numerical calculations, Fig.~\ref{alle2}(d) shows the conductance $G$ as a function of magnetic field for two different carrier densities representative for the low- and high-carrier-density regime. The high-carrier-density conductance ($V_{bg}$=~-15.6~V) shows the oscillating behavior as described before, the low-density trace ($V_{bg}$=~-6.2~V) exhibits first a conductance increase followed by a conductance decrease. In the following, we discuss the observed behavior with the help of numerical simulations.

\section{Numerical Transport Simulations}

The experimental data in Fig.~\ref{alle2} will give us important insight into the nature of the defects relevant in our GNRs. Specifically, in this section we will focus on the visibility of the SdH oscillations, the positive magnetoconductance at low carrier densities and fields up to about 20~T, and the rather high zero-field resistance at both low and high carrier densities. To this end, we have performed numerical magnetotransport simulations of (armchair) graphene nanoribbons with realistic sizes (\mbox{$L=320\,$nm}, \mbox{$W\sim 25\,$nm}). Since Ohmic scaling is not applicable at those length scales~\cite{Xu2011} we do not expect a full quantitative match between theory and experiment. However, the qualitative behavior will be well reproduced by the simulations since the system size is of the same order as the experimental samples. We used the well-known graphene tight-binding Hamiltonian in nearest neighbor (n.n.) approximation,
\begin{equation}
\label{eq:H}
 H = \sum_{ i,j\, \text{n.n.} } t_{ij} c_i^\dagger c_j ,
\end{equation}
where for finite magnetic field the corresponding hopping integral is given by $t_{ij} = - t \exp[i e/\hbar \int_{\boldsymbol{x}_i}^{\boldsymbol{x}_j} d\boldsymbol{s} \boldsymbol{A}(\boldsymbol{x}) ]$, with constant $t\approx$~2.7~eV and the vector potential $\boldsymbol{A}(\boldsymbol{x})$. The conductance was then computed using an adaptive recursive Green-function method, capable of treating arbitrarily shaped systems \cite{Wimmer2009}.

\begin{figure}
\includegraphics[width=8.4cm]{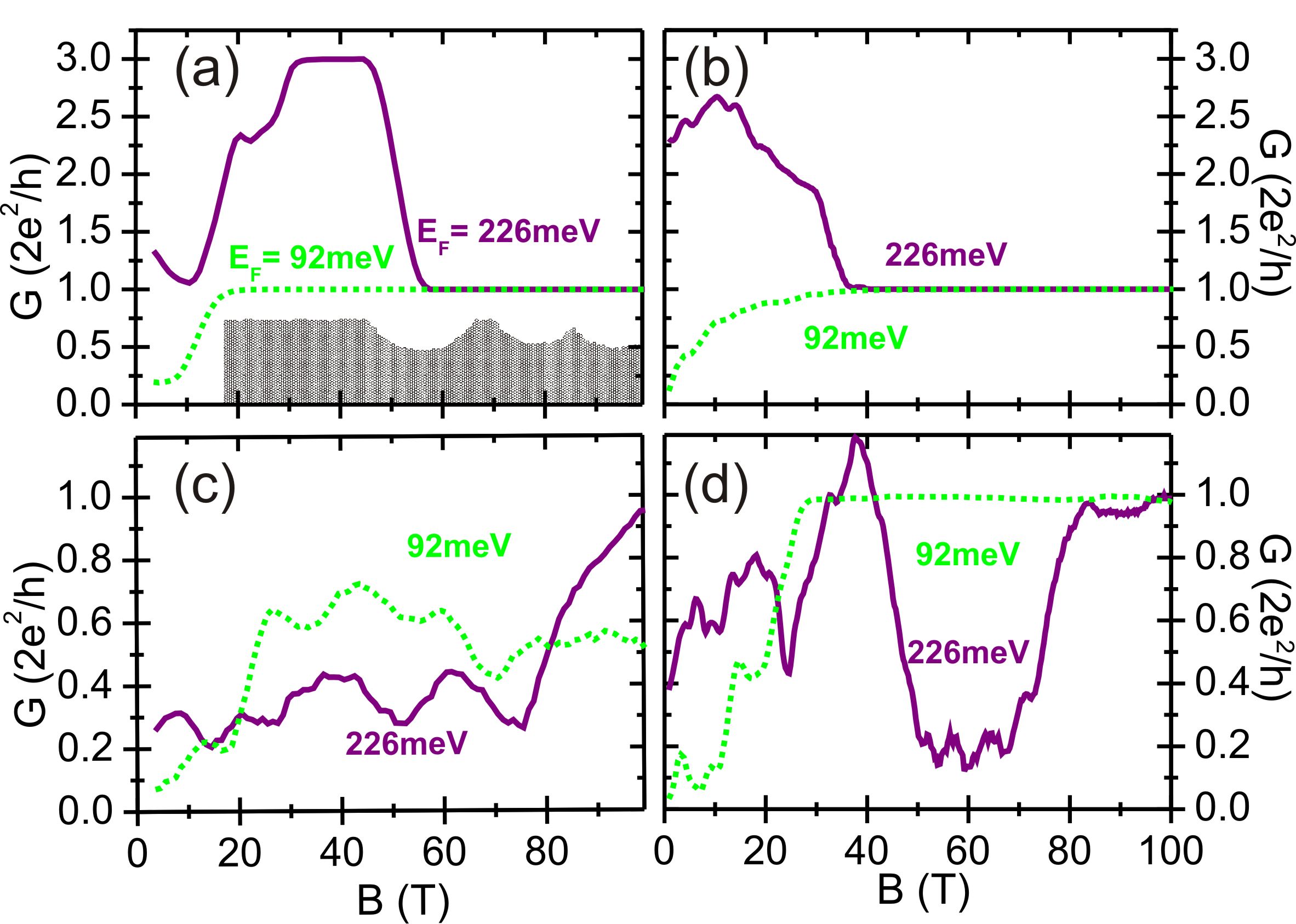}
\caption{(Color online) Magnetoconductance of armchair GNRs (\mbox{$L=320\,$nm}, \mbox{$W\sim
25\,$nm}) calculated numerically, using tight-binding simulations
\cite{Wimmer2009} and different disorder models. (a) Edge disorder (cf. text, inset: a close up of the ribbon edge with disorder). (b) Long-range Gaussian disorder (puddles,
cf. text). (c) Short-range impurities. We used Gaussian disorder with a
decay length of $\sim 0.44\,$nm. The height of the individual Gaussian
potentials is randomly distributed within the interval $[-\delta,
\delta]$ with $\delta= 0.1\,t$ and the impurity density is $p=15$\%.
(d) Edge disorder and short-range Gaussian disorder. Here $\delta=
0.09\,t$ and $p=8$\%.
}
\label{Fig:figure8}
\end{figure}

To appropriately describe the experimental situation, we considered different types of disorder. Since the fabrication process certainly leads to disordered edges, we took this into account also in the numerical simulations. To this end, we cut `chunks' of about $4\,$nm out of the graphene lattice at random positions close to the edge, which simulates the large-scale edge roughness that occurs due to e-beam resist roughness and the random nature of reactive ion etching. Additionally, we accounted for edge roughness on a smaller scale of a few lattice constants using a model introduced in Ref.\,\cite{Mucciolo2009}: About $10$ percent of the edge atoms are randomly removed and subsequently dangling bonds are additionally removed. This procedure was repeated $5$ times to yield an edge roughness of a few lattice constants. The numerical results, however, showed that both types of disorder yield similar results. In the following, in the case of edge disorder, both mechanisms will always be included.

In addition to the edge disorder, we studied two types of bulk potential disorder. On the one hand, we modeled so-called electron-hole puddles, {\em i.\,e.}, long range potential fluctuations due to charged impurities trapped beneath the graphene ribbon in the silicon-oxide substrate. Second, we also consider shorter-ranged impurity potentials, that can arise due to adsorbates, defects or charged impurities. In both cases, we add Gaussian on-site potentials to the tight-binding Hamiltonian (\ref{eq:H}). For the puddles, we use Gaussians with a decay length of $\sim 8.5\,$nm and a total height of $\sim 80\,$meV, which is comparable to the experimentally determined values \cite{Martin2008}. The impurities were modelled by Gaussians with a decay length of~$\sim 0.44\,$nm~\cite{CastellanosGomez2011}.

In Fig.~\ref{Fig:figure8}, we present our numerical results for magnetotransport through disordered nanoribbons at relatively high ($E_F \approx 226\,$meV) and lower ($E_F \approx 92\,$meV) carrier densities, corresponding to the Fermi energies of the experimental data in Fig.~\ref{alle2}(d).
First, we consider ribbons with edge disorder only [Fig.~\ref{Fig:figure8}(a)]. We find that while the zero-field conductance for low densities is comparable to the experiment, this is not the case for the high-density result. Upon increasing the field, the wavefunctions become more localized close to the edges. Without bulk disorder, backscattering is strongly suppressed, so that calculations yield nearly perfect quantum Hall plateaus for all densities already at moderate fields, in contrast to the experimental findings. This means that edge disorder alone cannot explain the experiment.
Considering only long-range Gaussian disorder [panel (b)], we find that the puddles are rather effective scatterers at low density, while they affect $G$ only little at high densities.
Simulations where only the short-range impurities are taken into account [panel (c)], show that indeed for strong enough scattering potentials, the zero-field conductance can be very close to the experimental data. However, such strong bulk disorder leads to backscattering even for very high magnetic field, so that at high carrier density no SdH oscillations can be observed.
This implies that indeed a combination of bulk \emph{and} edge disorder is necessary to describe the high-field experiments.
In panel~(d), we show the results for ribbons with disordered edges and short-range bulk disorder. In this case, the experimental findings for low and moderate field are reproduced semi-quantitatively. For low density, we find a strong increase of $G$ due to the formation of edge channels, while clear SdH oscillations are obtained at higher densities. The zero-field conductance fits well with the experiment. In contrast, in simulations that additionally include the long-range puddles, the difference in the zero-field conductance for high and low densities is much too high, thus we conclude that puddles are not the dominant scatterers in our samples. We note that beyond our disorder model interaction effects may further influence the measured conductance.

\section{High Field Insulating State at Low Densities}

We now turn our attention to the sample properties at high magnetic fields near the CNP.
As shown in Fig.~\ref{alle2}(a), the resistance at low temperatures initially decreases with $B$ and then diverges steeply by several orders of magnitude for $B > 20$~T.
While the initial negative magnetoresistance at low densities is explained in the previous section by the formation of edge channels related to the zero-energy Landau level (LL) in graphene, a crossover to a divergent resistance for $B > 20$~T requires another transport mechanism.
The zero-energy state in bulk graphene has been investigated by several research groups, and a strong increase in $R$ at the CNP and intense magnetic fields has been observed, resulting in a $B$-dependent LL splitting \cite{Giesbers2009, Zhang2010} and eventually a strongly insulating state~\cite{Checkelsky2008, Checkelsky2009}, the exact nature of which is still under debate \cite{DasSarma2011}.

\begin{figure}
\includegraphics[width=7.8cm]{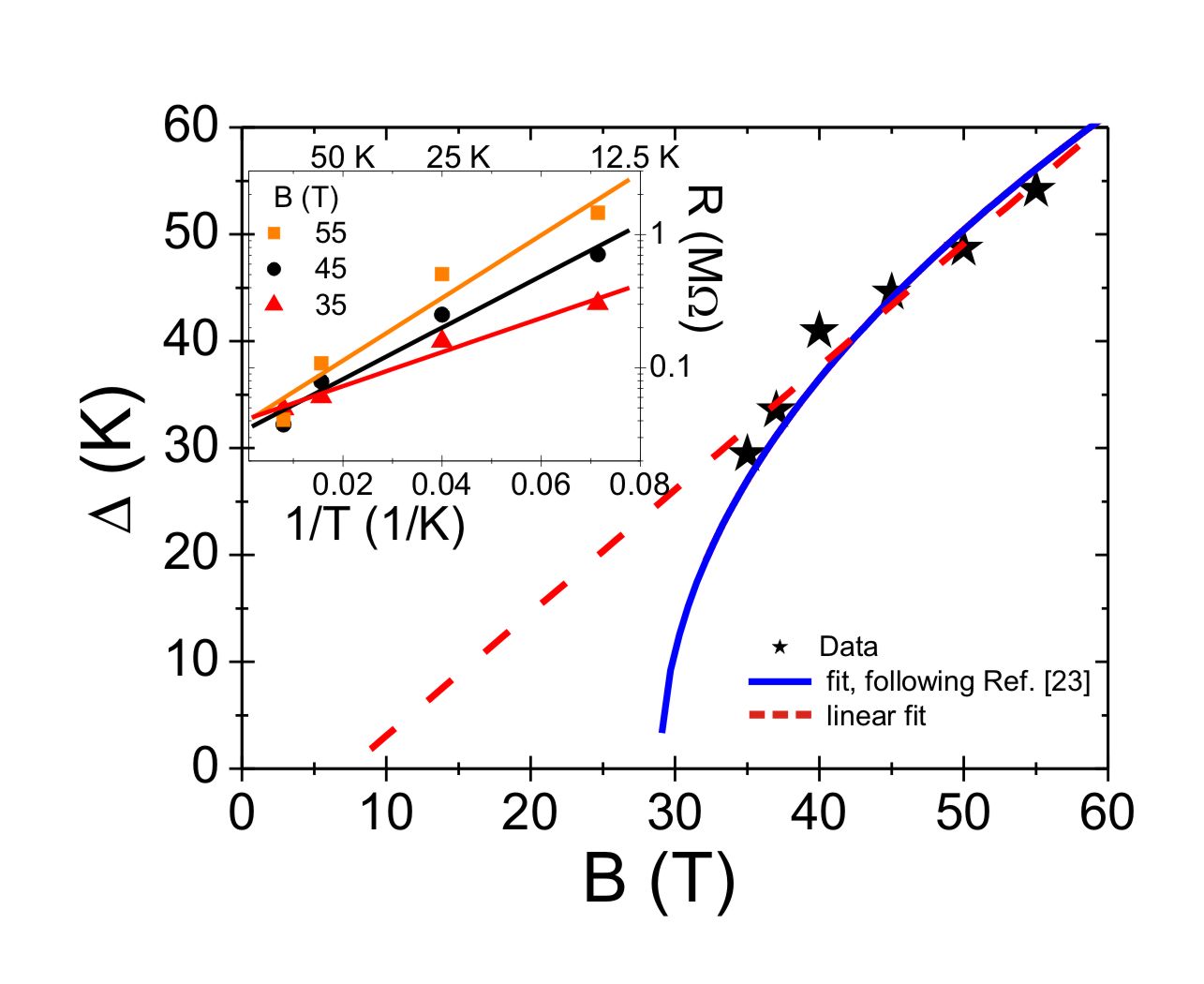}
\caption[gap]{(Color online) Energy gaps, $\Delta$, extracted from the slope of the Arrhenius plot for $T\geq$~14~K (inset). The (red) dotted line fits the Zeeman splitting, $\Delta = (g \mu_{B} B)/k_{B} - 8.9$~K,  with the Bohr magneton~$\mu_{B}$, the Boltzmann constant~$k_{B}$ and a gyromagnetic factor of $g$~=~1.73. The (blue) continuous line is a fit following Ref.~\cite{Khveshchenko2001}, cf. text.
}
\label{gap}
\end{figure}

Adopting a simple model involving the opening of a field-dependent spin gap \cite{Giesbers2009},
we can fit the temperature dependence of $R$ for $T\geq$~14~K in an Arrhenius plot for distinct magnetic-field values (inset of Fig.~\ref{gap}).
In Fig.~\ref{gap}, energy gaps, $\Delta$, are extracted from linear fits to the Arrhenius plot.
The gap $\Delta$ shows a linear dependence on $B$ (Fig.~\ref{gap}), consistent with spin splitting of the zero-energy LL, with the gyromagnetic factor $g$=1.73. However, another origin of the gap can also be considered. Following for example Ref.~\cite{Khveshchenko2001}, we can fit $\Delta \propto C \cdot (B - B_{c})^{0.5}$ with $B_{c}\approx$~29~T and $C\approx$~11, see Fig.~\ref{gap}, suggesting a chiral symmetry breaking transition. Comparing these different models we conclude that both mechanisms are compatible with our data, but the exact nature of the gap cannot be determined experimentally. For lower temperatures ($T \leq 7$~K), however, the resistance diverges strongly with $B$, and a simple activated behavior can no longer explain our data.
This divergent behavior of $R$ in our GNRs resembles a field-induced transition to a strongly insulating state reported in bulk graphene at low $T$ \cite{Checkelsky2008,Checkelsky2009}. In cleaner samples the transition to the insulating state occured at significantly lower fields.

Given the sample geometry displayed in Fig.~\ref{GNR}(a), we note that (bulk) graphene leads are attached to the GNR.
Since our GNRs, after patterning, have lower mobility than the bulk graphene leads the field required for the $B$-induced insulating state is expected to be also higher.
Therefore, the observed divergent $R$ at very high $B$ and low densities is tentatively attributed to the leads: when we apply high $B$-fields the leads become insulating
and mask the electron transport in the GNR.

\section{Conclusions}

In conclusion, we have performed transport experiments in graphene nanoribbons in pulsed high magnetic fields and corresponding transport simulations, based on a tight-binding model. This allows us to separate the contributions of different disorder types to magnetotransport. At least a combination of edge disorder and short-range bulk impurities is needed to reproduce the experimental results semi-quantitatively. The short-range bulk disorder is responsible for the partial suppression of the quantum Hall effect, while the edge disorder, together with the bulk disorder, provides sufficient backscattering to explain the observed high resistance at zero field for all carrier densities. Additionally, we observe a magnetic-field-induced insulating state at very low densities, which presumably originates from the bulk graphene leads.

\begin{acknowledgments}
We would like to thank B. Raquet for helpful discussions.
This research was supported by the Deutsche Forschungsgemeinschaft within GRK 1570 and by EuroMagNET under the EU Contract No. 228043.
\end{acknowledgments}


\begin{thebibliography}{20}
\bibitem[{FN3()}]{FN3}
\bibinfo{note}{This value does not change significantly if a contact resistance of up to 4~k$\Omega$ is taken into account. Our palladium contacts usually have a contact resistance of 1~k$\Omega$ or less.}
\bibitem[{FN1()}]{FN1}
\bibinfo{note}{Compared to Fig. \ref{GNR}(b) and Fig.~\ref{alle2}, these data
  were taken after thermal cycling where the CNP had shifted by about one Volt,
  but the mobility remained unchanged.}
\bibitem[{FN2()}]{FN2}
\bibinfo{note}{Here the resistance value exceeds the expected value of 4.3
  k$\Omega$ since it contains a series contribution of the Pd contacts and the
  bulk graphene leads, which are also in the quantum Hall regime.}
\end{thebibliography}
\end{document}